\documentclass[runningheads]{llncs}
\usepackage{graphicx}
\begin{document}
\title{Temporal EigenPAC for dyslexia diagnosis}
%
%
\author{Nicolás Gallego-Molina\inst{1} \and
Marco Formoso\inst{1} \and
Andrés Ortiz\inst{1,3} \and
Francisco J. Martínez-Murcia\inst{2,3}\and
Juan L. Luque\inst{4}}
\authorrunning{Nicolás Gallego-Molina et. al}
%
\institute{Department of Communications Engineering, University of Malaga (Spain) \and Department of Signal Theory, Telematic and Communications, University of Granada (Spain) \and Andalusian Data Science and Computational Intelligence Institute (DasCI) \and Department of Developmental Psychology, University of Malaga (Spain)
\email{njgm@ic.uma.es}}
\maketitle           
\begin{abstract}
Electroencephalography signals allow to explore the functional activity of the brain cortex in a non-invasive way. However, the analysis of these signals is not straightforward due to the presence of different artifacts and the very low signal-to-noise ratio. Cross-Frequency Coupling (CFC) methods provide a way to extract information from EEG, related to the synchronization among frequency bands. However, CFC methods are usually applied in a local way, computing the interaction between phase and amplitude at the same electrode.
In this work we show a method to compute PAC features among electrodes to study the functional connectivity. Moreover, this has been applied jointly with Principal Component Analysis to explore patterns related to Dyslexia in 7-years-old children. The developed methodology reveals the temporal evolution of PAC-based connectivity. Directions of greatest variance computed by PCA are called eigenPACs here, since they resemble the classical \textit{eigenfaces} representation. The projection of PAC data onto the eigenPACs provide a set of features that has demonstrates their discriminative capability, specifically in the Beta-Gamma bands.

\end{abstract}
\section{Introduction}
Developmental Dyslexia (DD) is one learning disability disorders with a higher prevalence, affecting between 5\% and 13\% of the population ~\cite{Peterson2012}. It has an important social impact causing effects in children like low self-esteem and depression and may be a cause for school failure.

The diagnostic of DD is an important issue for procure the intervention programs that help to adapt the learning process for dyslexic children. In this way, an early diagnosis is essential, which has historically been a complex task due to the use of behavioural tests. These tests depend on the motivation of each children and also have the inconvenient of include writing and reading tasks which postpone the start of diagnosis (i.e. it is not possible to diagnose pre-readers).

This aspect is changing with the use of biomedical signals, which provide objective and quantifiable measures to study the neural basis of the healthy brain and its pathologies. A relevant method to obtain information of brain activity is the electroencephalography (EEG), which allows to acquire brain signals in a non-invasive way. This technique can be used to quantify the functional activity of the brain while developing a specific task.
In particular, EEG signals have been used to explore the neurological origin of DD in ~\cite{Power2016,Ortiz2020,Ortiz2019}, towards the advance in the knowledge of dyslexia and its objective diagnosis. 

One way to build functional models that help to understand the brain processes developed while the subject is developing a specific task, is through connectivity. In other words, it consists in measuring how the different brain areas cooperates in any manner while processing information. On the other hand, neural oscillations are produced mainly in five frequency bands: Delta (0.5-4) Hz, Theta (4-8) Hz, Alpha (8-12) Hz, Beta (12-30) Hz and Gamma ($>$ 30 Hz). The exploration of the relationship among these bands has demonstrated to provide useful information to characterize the brain activity. This way, Cross Frequency Coupling  (CFC) is a technique to explore interactions and (also called couplings) between frequency bands and has undergone and special attention in recent years. 

In the present work, EEG signals are used to explore the functional connectivity. Specifically, the Phase Amplitude Coupling (PAC), a type of CFC, is calculated to analyze and identify temporal patterns in dyslexic and non-dyslexic subjects. Then, Principal Component Analysis (PCA) is used for identify and extract patterns in order to perform a classification using SVM for a differential diagnosis. 

The paper is organized as follows. Section 2 presents details of the database and describes the auditory stimulus and the methods used. Then, Section 3 presents and discusses the classification results, and finally, Section 5 draws the main conclusions and the future work.

\section{Materials and Methods}

\subsection{Database and stimulus}
The EEG data used in this work was provided by the Leeduca Study Group at the University of Málaga ~\cite{Ortiz2020B}. EEG signals were recorded using the Brainvision acticHamp Plus with 32 active electrodes (actiCAP, Brain Products GmbH, Germany) at a sampling rate of 500 Hz during 15 minutes sessions, while presenting an auditory stimulus to the subject. A session consisted of a sequence of white noise stimuli modulated in amplitudes at rates 2, 8, and 20 Hz presented sequentially for 5 minutes each.

The present experiment was carried out with the understanding and written consent of each child’s legal guardian and in the presence thereof. Forty-eight participants took part in the present study, including 32 skilled readers (17 males) and 16 dyslexic readers (7 males) matched in age (t(1) = -1.4, p $>$ 0.05, age range: 88-100 months). The mean age of the control group was 94, 1 $\pm$  3.3 months, and 95, 6 $\pm$ 2.9 months for the dyslexic group. All participants were
right-handed Spanish native speakers with no hearing impairments and normal or corrected–to–normal vision. Dyslexic children in this study have all received a formal diagnosis of dyslexia in the school. None of the skilled readers reported reading or spelling difficulties or have received a previous formal diagnosis of dyslexia. The locations of 32 electrodes used in the experiments is in the 10–20 standardized system.

\subsection{Signal Prepocessing}
The EEG signals recorded were processed to remove artifacts related to eye blinking and impedance variation due to movements. It was used blind source separation with Independent Component Analysis (ICA) to remove artifacts corresponding to eye blinking signals in the EEG signals. Then, EEG signal of each channel was normalized independently to zero mean and unit variance and referenced to the signal of electrode Cz. Baseline correction was also applied. Finally, the EEG signals were segmented into 15.02 s long windows in order to analyze PAC temporal patterns correctly ~\cite{Dvorak2014}. This is the minimum appropriate window length for which a sufficiently high number of slow oscillation cycles are analyzed, shorter windows lead to overestimates of coupling and lower significance. This adequate window length is one of the main requisites for robust PAC estimation and appropriate statistical validation of the result by surrogate tests without using long data windows that assumes stationarity of the signals within the window.

\subsection{Phase-Amplitude Coupling (PAC)}
Cross-frequency coupling(CFC) has been proposed to coordinate neural dynamics across spatial and temporal scales~\cite{Aru2015}, it serve as a mechanism to transfer information from large scale brain networks and has a potential relevance for understanding healthy and pathological brain function. In particular, Phase-Amplitude Coupling has received significant attention ~\cite{Dvorak2014,Meij2012} and may play an important functional role in local computation and long-range communication in large-scale brain networks~\cite{Canolty2010}. 

PAC describes the coupling between the phase of a slower oscillation and the amplitude of a faster oscillation. Concretely, in this work we explore the modulation of the amplitude of the Gamma (30–100) Hz frequency band by the phase of the Delta (0.5-4) Hz, Theta (4-8) Hz, Alpha(8-12) Hz and Beta(12-25) Hz bands. There are different PAC descriptors for measuring PAC ~\cite{Tabassum2019}. In this work, we use the Modulation Index (MI)~\cite{Tort2008,Tort2010}, although there is no convention yet of how to calculate phase-amplitude coupling and much heterogeneity of phase-amplitude calculation methods used in the literature ~\cite{Hulsemann2019}. 

For calculating MI as in~\cite{Tort2010}, first all phases are binned into eighteen 20 degrees intervals and the average amplitude of the amplitude-providing frequency in each phase bin of the phase-providing frequency is computed and normalized by the following formula:
\begin{equation}
p(j)= \frac{\bar{a}}{\sum^{N}_{j=1}\bar{a}_k}
\end{equation}
where $\bar{a}$  is  the  average  amplitude  of  one  bin,  $k$  is  the  running index  for  the  bins,  and  $N=18$  is  the  total  amount  of  bins; $p$  is  a vector  of $N$ values. 

Then, Shannon entropy \emph{H(p)} is computed by means of

\begin{equation}
\label{eq_shannon}
H(p)=-\sum^{N}_{j=1}P(j)log{P(j)}
\end{equation}

where  $p$  is  the  vector  of  normalized  averaged  amplitudes  per phase bin. This represents  the  inherent  amount  of  information  of  a  variable. If the Shannon  entropy  is  maximal, all the phase bins present the same amplitude (uniform distribution). Thus, the existence of phase-amplitude coupling is characterized by a deviation of the amplitude distribution from the uniform distribution in a phase-amplitude plot. To measure this, the Kullback–Leibler (KL) distance of a discrete distribution $P$ from a distribution $Q$ is used and it is defined as 

\begin{equation}
KL(P,Q)=\sum^{N}_{j=1}P(j)log\frac{P(j)}{Q(j)}
\end{equation}

and the KL distance is relate to the Shannon entropy by the following formula

\begin{equation}
KL(U,P)=logN-H(p)
\end{equation}

where U is the uniform distribution and P is the amplitude distribution defined earlier by p(j) **? Finally, the  raw  MI  is  calculated  by  the following formula:

\begin{equation}
MI= \frac{KL(U,X)}{logN}
\end{equation}

In this work, PAC is measured by MI in each data segment, enabling  the exploration of the temporal evolution of the response to specific auditory stimuli. This is achieved with the use of Tensorpac ~\cite{tensorpac2020}, an open-source Python toolbox dedicated to PAC analysis of neurophysiological data. Tensorpac provides a set of efficient methods and functions to implements the most common PAC estimation methods, such as the Modulation Index used in the present work.

\subsection{Dimensionality Reduction and Classification}
Principal Component Analysis (PCA) is a widely used method to perform dimensionality reduction. It is a well known multivariate analysis technique used in many studies ~\cite{Abdulhamit2010,Markiewicz2009} to significantly reduce the original high-dimensional feature space to a lower-dimensional subspace spanned by a number (n) of Principal Components (PC), while preserving the variation of the dataset in the original space as much as possible. 

A well known application of PCA the so-called eigenimage decomposition, which results from the application of PCA to images. This is used in different works such as ~\cite{Illan2011,Alvarez} and adapted from the eigenface approach of Turk and Pentland ~\cite{Turk1991}. In this work this approach is used allowing to detect underlying patterns that differentiate individuals within a population, even if the differences are subtle. 

Here, PCA is applied in the time axis to compute the maximum variance directions of the PAC along the EEG segments. Thus, we obtain the PCs or eigenvectors of the covariance  matrix of the a dataset composed by N vectors, corresponding to the MI values of each 31 electrodes for the ten segments of every subject. These eigenvectors describe a set of features that  characterize the variation between the PAC measured in each temporal segment. As usual, they are sorted in decreasing  explained variance order. We can display these eigenvectors in topoplots, representing the principal components at each electrode position. In order to keep the traditional notation, we called these PC as \textit{eigenPACs}. Then, we selected the eigenPACs that have the largest eigenvalues which therefore account for the most variance within the set of PAC matrix, composing a M-dimensional subspace.

For the sake of clarity, let the measured PAC vector set be $\Gamma_1,\Gamma_2,...,\Gamma_N$
of length equal to the number of electrodes. The average PAC of the dataset is defined as $\Gamma=\frac{1}{N}\sum^{N}_{n=1}\Gamma_n$. Each measured PAC differs from the average by the vector $\Phi_i=\Gamma_i-\Gamma$ with $i=1,2,...,N$. On this set, a PCA transformation is applied  obtaining $M$ orthogonal vectors $u_i$ which best describes the distribution of the data. This vectors satisfy that
\begin{equation}
\lambda_1= \frac{1}{N}\sum^{N}_{n=1}(u^T_i\Phi_n)^2
\end{equation}
is maximum, subject to 
\begin{equation}
u^{T}_{i}u_{j}= \delta_{ij}
\end{equation}
where $\delta_{ij}$ is the Kronecker delta, and $u_{i}$ and $\lambda_i$ are the eigenvectors and eigenvalues, respectively, of the covariance matrix:
\begin{equation}
C= \frac{1}{N}\sum^{N}_{n=1}\Phi_n\Phi^T_i=AA^T
\end{equation}
where the matrix $A={\Phi_1,...,\Phi_N}$. These eigenvectors are what we refer as eigenPAC. Usually, the first few eigenPACs explain almost the whole variance, so only a number $M'<M$ is necessary to appropriately describe the dataset ~\cite{Alvarez}. Thus, the computacional complexity of the diagonalization process to obtain the eigenPAC basis is significantly reduced.

In essence, the eigenPACs define a new space in which each component explains the maximum variance in the data represented by its eigenvalue and its correlation is minimized. The projection of the PAC vectors onto the eigenPAC space, will determine the coordinates of each PAC vector in this subspace. It is expected that this
projection produce a pattern more suitable for class separation than the projection onto the average PAC space, due to the eigenPAC decorrelation ~\cite{Illan2011}.

For the classification we use this projections of the data on the new  basis to train a Support Vector Machine (SVM). Hence, the PAC vectors measured for the ten segments EEG signals are projected on the eigenPAC basis obtaining $N$ vectors each of them with its corresponding class label defined as control and dyslexic. Then, using the training data the SVM separates this set of binary labelled data with a hyperplane that is maximally distant from the two classes~\cite{Alvarez}.

\section{Experimental Results}

In this section, we show the experimental results obtained in with the PAC analysis, eigenPAC representation and classification. As mentioned before, the EEG data was segmented into ten temporal windows of 15.02 s for the analysis of the temporal evolution of the response to the stimulus. This analysis was performed with the measure of PAC over each segment.

\subsubsection{PAC Results} 
\begin{figure}[!htb]
\includegraphics[width=\textwidth]{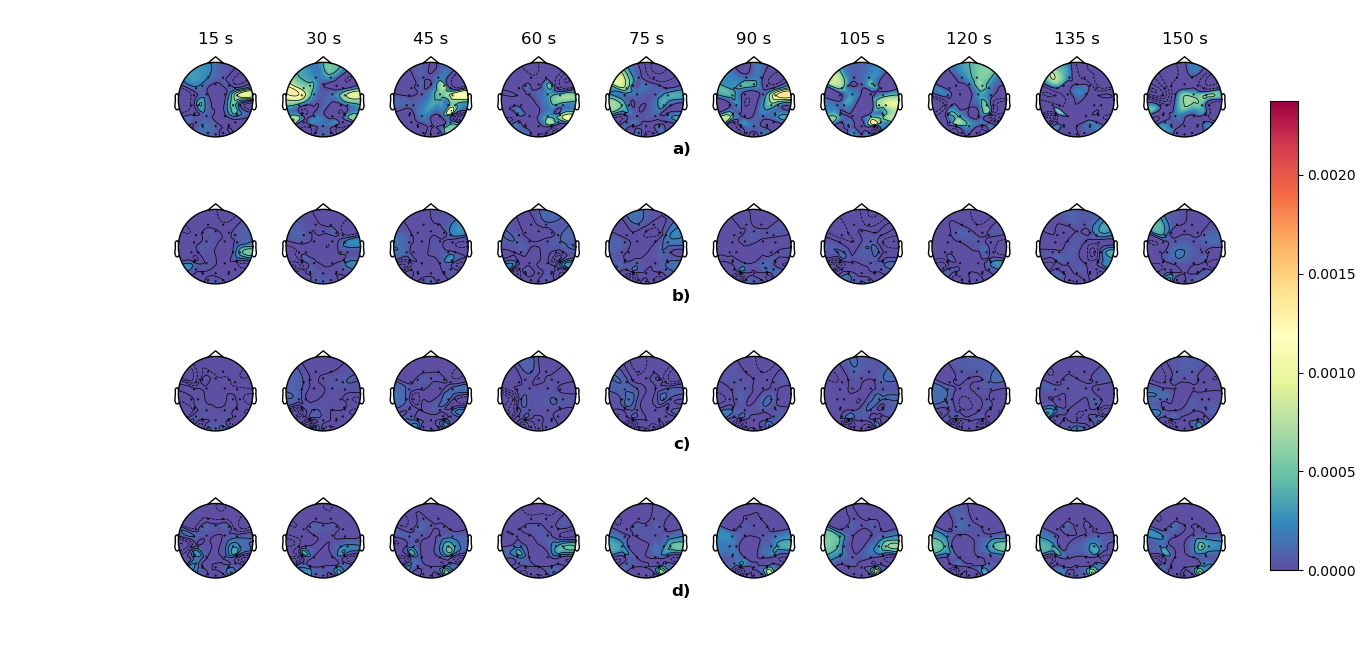}
\caption{Difference average MI topoplot for 2 Hz. a) Delta-Gamma b)Theta-Gamma c)Alpha-Gamma d)Beta-Gamma} \label{fig2}
\end{figure}

To analyze PAC connectivity we used the tensorpac tool ~\cite{tensorpac2020}. Thus,  we defined the frequency bands in which the PAC is measured (phase band and amplitude band) and we expected an identifiable temporal behaviour. This set of frequency band pairs are: Delta-Gamma, Theta-Gamma, Alpha-Gamma and Beta-Gamma. We measured the PAC for each subject and each temporal segment obtaining results for all the frequency bands. This results are represented with a set of ten topoplots showing the temporal evolution of the average PAC of dyslexic subjects and control subjects in each frequency pair. 
Figure.~\ref{fig2} shows the differences between the average MI value for the dyslexic group and the control group. In this Figure we represented each combination of frequency bands for which the PAC has been measured. These topoplots denote differences between the response of dyslexic and control subjects.

\subsubsection{EigenPAC Results}

\begin{figure}[!htb]
\includegraphics[width=\textwidth]{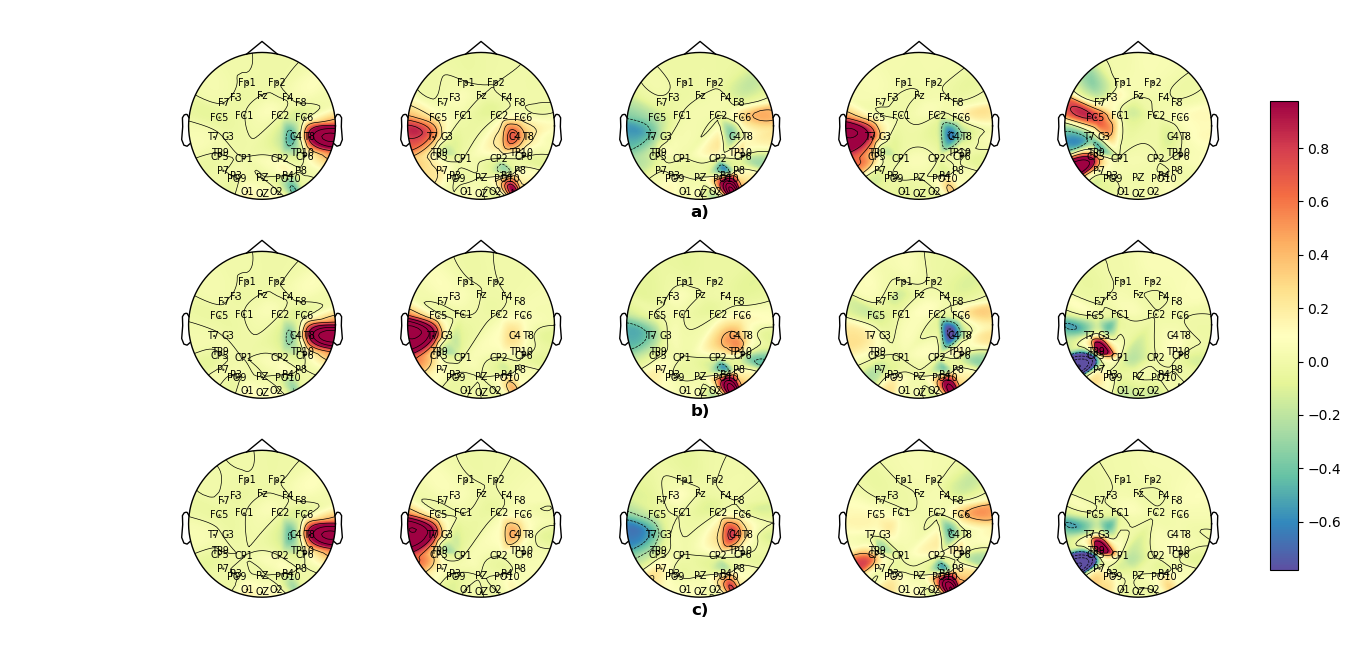}
\caption{First 5 eigenPAC for Beta-Gamma. a)2 Hz b) 8 Hz c) 20 Hz} \label{fig3}
\end{figure}

PCA has been applied in two different ways. In the first case, PAC features from all the subjects have been used to obtain the PCs, as in the case of \textit{eigenfaces} problem. This aims to obtain a representation of the overall database in terms of the maximum variance directions. To carry out this experiment, a matrix is created containing the MI value computed from temporal segments of all subjects. Specifically, this matrix contains N*(number of segments) rows corresponding to the number of subjects multiplied by the number of segments and M columns corresponding to MI of each of the 31 electrodes. Then, PCA is applied and we obtain a set of PCs of which the first five represent the most part of the variance.  In Fig.~\ref{fig3} we can see the representation of the eigenPAC for the first 5 PC indicating the area where there is a major temporal variation in the measured MI for the Beta-Gamma PAC. This eigenPAC are different for each stimulus and the first topoplot describes the maximal data variation. 

\begin{figure}[!htb]
\includegraphics[width=\textwidth]{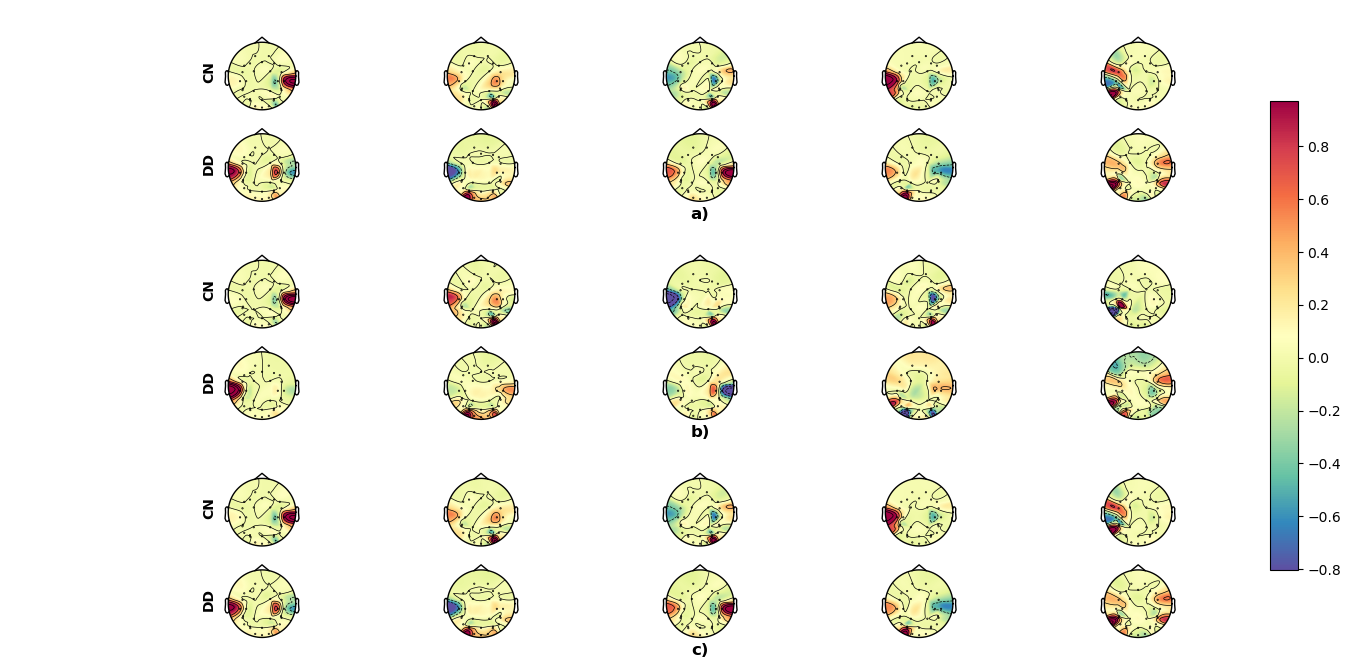}
\caption{First 5 eigenPAC. a)2 Hz b) 8 Hz c) 20 Hz} \label{fig4}
\end{figure}

A second experiment is performed by applying PCA to dyslexic and control groups separately. This results similar as the previous case, but only with subjects of one group. Thus, we achieve a better representation of the temporal variation for each group, obtaining a set of PCs describing the maximal variation for the dyslexic subjects and another set of PCs for the control subjects. These eigenPAC, represented in Fig.~\ref{fig4}, show the variation specifically related to the temporal response to the auditory stimulus. As shown, we found global similarities for each group. Furthermore, this helps to identify the characteristic patterns of each class that are used by classifications algorithms reaching a better performance.  

In Fig.~\ref{fig4}, for each case there are represented in the upper row the topoplots containing the information related to control eigenPAC and in the row below the dyslexic eigenPAC.

\subsubsection{Classification}

Once the application cases of PCA are defined, we used the resulting eigenPAC to train a SVM classifier. Therefore, there are two ways of perform a classification depending on how we compute the PCs that form the eigenPAC. These are used to project the data into the lower-dimensional space to obtain the feature space of the SVM classifier.

In the first case, the PCs are obtained from the application of PCA over all the subjects, dyslexic and control. Then, this PCs correspond to the eigenPAC that we use to project the data for the SVM. In this part, a k-fold stratified cross validation scheme is employed to separate the data into train sets and test set, specifically, 5-fold cross-validation is used. 

In the second case, the application of PCA separately over the two classes provide a set of PCs for each class. Then each subject data is projected onto the control and DD components, and these projections are concatenated to compose the feature vector. The process of cross validation is the same as in the above case just with a different set of PCs.

\begin{figure}[!htb]
\centering
\includegraphics[width=0.7\textwidth]{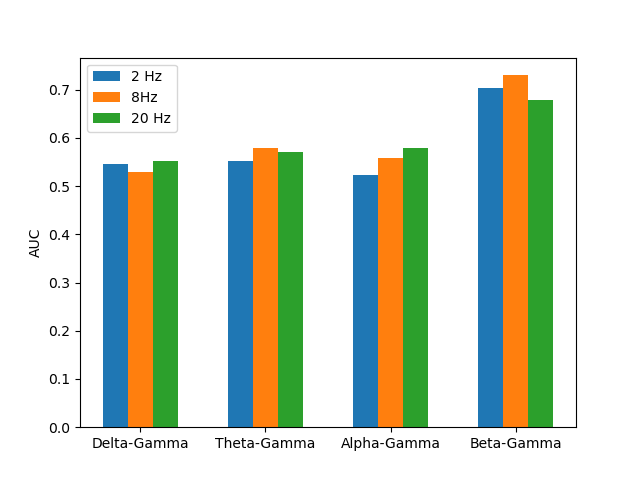}
\caption{Max AUC for each band combination and stimulus} \label{fig5}
\end{figure}

The metrics used in the classification are the accuracy, sensitivity, specificity and Area Under ROC curve (AUC). The evaluation metrics are described in Table~\ref{tab1}. The results show the performance of the two classification scenarios for the best case, which corresponds to the Beta-Gamma PAC as represented in Fig.~\ref{fig5}. We can see that there is a improvement related with the application of PCA separately for the two classes.

\begin{table}[!htb]
\centering
\caption{Classification comparative using the Beta-Gamma PAC}\label{tab1}
\begin{tabular}{|l|c|c|c|c|c|c|}
\hline
Metrics & \multicolumn{3}{|c|}{ PCA with all the subjects} & \multicolumn{3}{|c|}{PCA with each class separately}\\
\hline
& 2 Hz & 8 Hz & 20 Hz & 2 Hz & 8 Hz & 20 Hz\\
\hline
Accuracy &  0.572 & 0.611 & 0.561 & 0.654 & 0.653 & 0.594\\
Sensitivity &  0.501 & 0.551 & 0.515 & 0.651 & 0.635 & 0.551\\
Specificity &  0.743 & 0.757 & 0.675 & 0.661 & 0.696 & 0.7\\
AUC &  0.65 & 0.705 & 0.622 & \bf{0.699} & \bf{0.721} & \bf{0.65}\\
\hline
\end{tabular}
\end{table}

Therefore, the second case generate higher metrics achieving a better classification performance with a greater AUC and accuracy for 2 Hz, 8 Hz and 20 Hz. We present the results for this case in Fig.~\ref{fig5} where the max AUC is represented for each band combination and each stimulus. Showing that in the Beta-Gamma there are temporal pattern that are distinctive of each class.

\begin{figure}[!htb]
\centering
\includegraphics[width=0.7\textwidth]{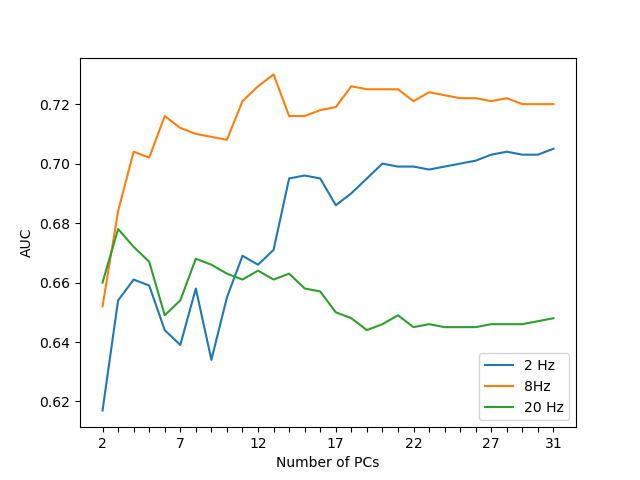}
\caption{Max AUC for each number of PCs} \label{fig6}
\end{figure}

In the case of Beta-Gamma we perform a PC number sweep obtaining the explained variance and the max AUC for the classification with each number of components in Fig.~\ref{fig6}

\section{Conclusions and Future Work}

In this work we present a classification method for EEG signals based on the study of functional connectivity PAC and the use of eigenPAC resulting of applying PCA to train a SVM classifier. The concept of eigenPAC helps to extract the underlying pattern that differentiates the temporal response between control and dyslexic subjects. Also, it can be represented in topographic plots to visualize the areas with a greater variation principally corresponding to the temporal evolution.

The classification results suggest differential patterns in the Beta-Gamma bands that allows to discriminate between control and dyslexic subjects, obtaining the highest AUC for the 8 Hz stimulus. This points the bands in which the response to the stimulus has differences across the temporal segments and encourage continuing the approach shown in this work. As future work, it is interesting to study the effect of using other windows length and to improve the necessary PAC analysis with the use of decomposition methods such as MEMD to accurately extracts the oscillatory components of the EEG signals.

\section*{Acknowledments}
This work was partly supported by the MINECO/FEDER under PGC2018-098813-B-C32 project. We gratefully acknowledge the support of NVIDIA Corporation with the donation of one of the GPUs used for this research. Work by F.J.M.M. was supported by the MICINN ``Juan de la Cierva - Formaci\'on'' Fellowship. We also thank the \textit{Leeduca} research group and Junta de Andalucía for the data supplied and the support.

%
%
%
%


\end{document}